\documentclass[pre,twocolumn,aps,superscriptaddress,showpacs,floatfix]{revtex4}

\usepackage{amsmath}
\usepackage{amssymb}
\usepackage{amsbsy}
\usepackage{graphicx}
\usepackage{color}

\begin{document}
 
\title{Nonlinear response of inertial tracers in steady laminar flows:\\ differential and absolute negative mobility}

\author{A. Sarracino}
\affiliation{CNR-ISC and Dipartimento di Fisica, Sapienza Universit\`a di Roma, p.le A. Moro
2, 00185 Roma, Italy}

\author{F. Cecconi}
\affiliation{CNR-ISC and Dipartimento di Fisica, Sapienza Universit\`a di Roma, p.le A. Moro
2, 00185 Roma, Italy}

\author{A. Puglisi}
\affiliation{CNR-ISC and Dipartimento di Fisica, Sapienza Universit\`a di Roma, p.le A. Moro
2, 00185 Roma, Italy}

\author{A. Vulpiani}
\affiliation{Dipartimento di Fisica, Sapienza Universit\`a di Roma, p.le A. Moro
2, 00185 Roma, Italy}

\begin{abstract}
We study the mobility and the diffusion coefficient of an inertial
tracer advected by a two-dimensional incompressible laminar flow, in
the presence of thermal noise and under the action of an external
force. We show, with extensive numerical simulations, that the
force-velocity relation for the tracer, in the nonlinear regime,
displays complex and rich behaviors, including negative differential
and absolute mobility.  
These effects rely upon a subtle coupling
between inertia and applied force which induce the tracer to persist in particular regions of phase space with a velocity opposite to the force.
The relevance of this coupling is revisited in the framework of
non-equilibrium response theory, applying a generalized Einstein
relation to our system.
The possibility of experimental observation of these results is
also discussed.
\end{abstract}


\maketitle

\emph{Introduction.}-- Understanding the response of a system to an
external stimulus from the observation of the unperturbed dynamics,
represents a central issue in statistical mechanics. For weak
perturbations of an equilibrium state, the fluctuation-dissipation
theorem (FDT) solves the problem, expressing the system response in
terms of correlation functions~\cite{kubo}. Generalizations of this
result have been recently derived, to address the much more complex
issue of predicting the response in nonequilibrium
conditions~\cite{CR03,CLZ07,MPRV08,seifert,BM13}, when detailed
balance does not hold and currents cross the system, or in the
nonlinear response regime, where higher order response functions have
to be taken into account~\cite{BB05,LCSZ08,D12}. All these
approaches point out the role played by the coupling among degrees of
freedom which emerges out of equilibrium, adding extra-terms to the
standard FDT~\cite{SS06,VBPV09,SVGP10,C11,GPSV14}.

A paradigmatic problem within such a nonlinear response theory is
concerned with the dynamics of a tracer particle, traveling in a
complex medium under the action of an external field $F$. In
particular, one is interested in the force-velocity relation $v(F)$,
or the mobility $\mu(F)=v(F)/F$, and the diffusion coefficient $D(F)$
of the tracer particle. These curves can be strongly affected by the
interaction between the tracer and the surrounding medium, and can
show striking nonlinear behaviors.  This kind of problem has
originated in the field of active microrheology of complex fluids,
such as emulsions, suspensions, polymer, and micellar
solutions~\cite{SM09,PV14}, where information on the structure of the
host medium is inferred from the motion of a biased probe embedded in
it. In this context inertia is usually negligible, whereas the 
force-velocity relation in non-overdamped systems, which play an important
role in fluid dynamics~\cite{TB09}, seems much less studied.

One of the surprising effects observed in the force-velocity relation
of several models of biased tracers in nonequilibrium systems is a
negative differential mobility (NDM). This means that the tracer
velocity, after increasing linearly according to linear response,
displays a nonmonotonic behavior, characterized by a maximum for a
certain value of the external driving field. Just beyond this value,
the differential mobility $d\mu/dF$ becomes negative, implying a
slowing down of the particle motion at increasing force.  This kind of
phenomenon, denoted with the telling expression ``getting more from
pushing less'', has been explained for nonequilibrium toy models
in~\cite{royce} and can be observed in different systems, such as
Brownian motors~\cite{CM96}, kinetically constrained models of glass
formers~\cite{JKGC08,S08} and driven lattice
gases~\cite{LF13,BM14,BIOSV14,BSV15,BIOSV16}, where analytic
approaches are possible~\cite{LF13,BIOSV14,BIOSV16}.  In most of the
aforementioned systems the non-linear behavior is due to a reciprocal
tracer-medium interaction, i.e. the tracer not only feels the action
of the solvent but influences it, modifying its microstructure.  More
generally, within the framework of nonequilibrium statistical
mechanics, the occurrence of NDM has been related to the concept of
dynamical activity of the tracer, which is a measure of time-symmetric
currents and expresses a ``jitteriness'' of the particle during its
motion~\cite{BBMS13,BM14,M16}.

Even more surprisingly, there exist cases of \emph{absolute} negative
mobility (ANM), $\mu<0$, where the particle travels against the external
force. This phenomenon can be realized in specific models, due the
carefully tuned coupling between colored noise, asymmetric spatial
structures, and driving field~\cite{RERDRA05,KMHLT06,MKTLH07,ERAR10}.

In this Letter we show that these kinds of behaviors can take place in
more realistic inertial tracer models, relevant in fluid dynamics.  In particular, we investigate
the linear and nonlinear response of an inertial particle moving in a
steady (incompressible) cellular velocity field, under the action of
an external force, and subject to thermal agitation. The presence of
inertia implies a non-trivial deviation of the particle's motion from
the trajectory of a fluid particle, typically leading to the
appearance of strongly inhomogeneous distributions -- a phenomenon
known as preferential concentration or particle
clustering~\cite{BBCLMT07,CCLT08}.  This can be responsible for an
enhanced probability of chemical, biological or physical interaction,
as for instance, for the time scales of rain~\cite{F02}, sedimentation
speed under gravity~\cite{LCLV08}, or the planetesimals formation in
the early Solar System~\cite{B99}.  Here we discover that such a
preferential concentration strongly depends upon the external force,
leading to a rich non-linear behavior for the average particle's
velocity, showing NDM and, in particular cases, even ANM. By an
  analysis of the tracer's trajectories we identify a possible
  mechanism responsible for such behaviors. Moreover we interpret our
  results within the framework of nonequilibrium response theory,
exploiting a generalized Einstein relation (GER), derived recently
in~\cite{BBMW10,BMW11}, which makes clear the role played by the
coupling between the velocity field and the tracer dynamics.

\emph{The model.}-- We consider the following equations of motion of
an inertial tracer particle in two dimensions, with spatial
coordinates $(x,y)$ and velocities $(v_x,v_y)$, subject to an external
force $F$ along the $x$ direction, and traveling through a
divergenceless cellular flow $(U_x,U_y)$
\begin{eqnarray}
\dot{x}&=&v_x, \qquad \dot{y}=v_y \label{eq1} \\ 
\dot{v}_x&=&-\frac{1}{\tau}(v_x-U_x)+F +\sqrt{2D_0}\xi_x \label{eq2} \\ 
\dot{v}_y&=&-\frac{1}{\tau}(v_y-U_y)+ \sqrt{2D_0}\xi_y \label{eq3} \\ 
U_x&=&\frac{\partial \psi(x,y)}{\partial y}, \qquad U_y=-\frac{\partial \psi(x,y)}{\partial x}. \label{eq4}
\label{model}
\end{eqnarray}
Here $\psi(x,y)=LU_0/(2\pi) \sin(2\pi x/L)\sin(2\pi y/L)$ is the
stream function and $\xi_x$ and $\xi_y$ are uncorrelated white noises
with zero mean and unitary variance. The velocity field here
  considered corresponds to two-dimensional convection and shows a
very rich behavior~\cite{YPP89,CMMV99}. In addition, it can be easily
realized in a laboratory, e.g. with rotating cylinders~\cite{SG88} or
in ion solutions in array of magnets~\cite{T02}.  Let us stress that
our system, even in the absence of external driving $F$, is out of
equilibrium because of the steady velocity field represented by the
non-gradient forces of Eq.~(\ref{eq4}). In what follows we measure
length and time in units of $L$ and $L/U_0$ respectively, setting
therefore $U_0=1$ and $L=1$, which defines a typical time scale of the
flow $\tau^*=L/U_0=1$. Another important ingredient of our model is
the presence of microscopic noise with molecular diffusivity $D_0$,
which guarantees ergodicity and is related to the temperature $T$ of
the environment by $D_0=T/\tau$~\cite{nota0}.  We stress that, in the
presence of an advection field, the statistic of the phase space
explored by the tracer, $\{x(t),y(t),v_x(t),v_y(t)\}$, even at $F=0$,
depends on both $\tau$ and $D_0$ in a nontrivial way, and the finite
value of $\tau$ has an important role for the concentration
properties~\cite{BBCLMT07}.  When $\tau \to 0$ (fluid particle limit,
where the tracer evolves according to the equation $\dot{\boldsymbol
  x}=\boldsymbol{U}+\sqrt{2 D}\boldsymbol{\xi}$ with $D$ an effective
diffusivity), because of $\boldsymbol{U}=(U_x,U_y)$ incompressibility,
the tracer visits the two-dimensional phase space in a uniform
way~\cite{nota}. The same happens for $\tau \to \infty$, when the
tracer is insensitive to the field and uniformly diffuses through the
flow.

\begin{figure}[!t]
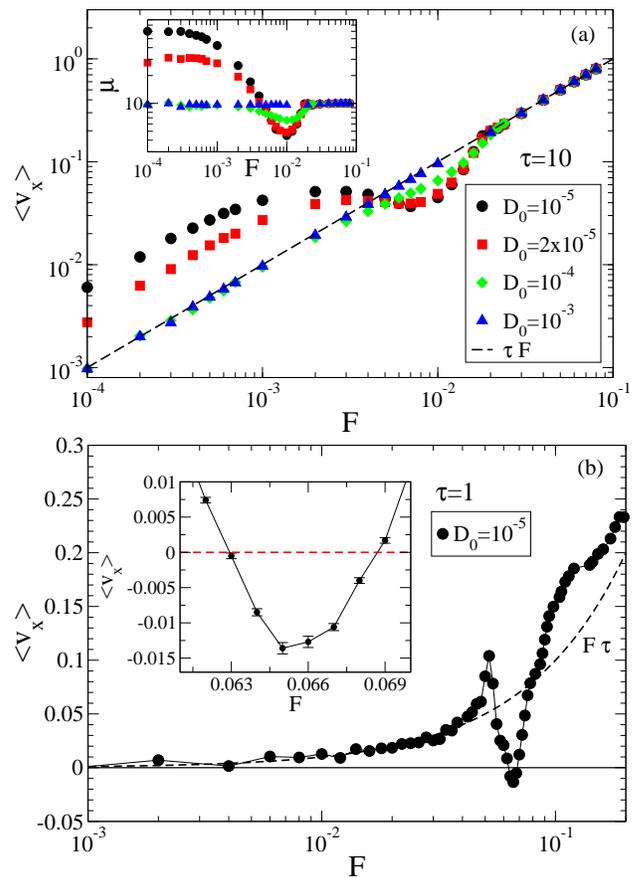

\includegraphics[width=.95\columnwidth,clip=true]{v_mu.eps}
\includegraphics[width=.95\columnwidth,clip=true]{tau1new.eps}
\caption{(a) Force-velocity relation $\langle v_x\rangle(F)$ and mobility $\mu$ (inset) for
 different values of $D_0$, in the case $\tau=10$. NDM is observable
 for $D_0=10^{-5}$ and $D_0=2\cdot 10^{-5}$, around $F\sim 4\cdot
 10^{-3}$. (b) Force-velocity relation $\langle v_x\rangle (F)$ in the case $\tau=1$ for 
  $D_0 =10^{-5}$. Notice the negative peak, corresponding to ANM, observed in a range of forces near $F\sim 
  6.5\cdot 10^{-2}$, which is magnified in the inset. }
\label{fig:tau10}
\end{figure}

\emph{Negative differential and absolute mobility.}-- Here we are mainly interested in the behavior of the stationary
velocity $\langle v_x \rangle = \tau F + \langle U_x[x(t),y(t)]
\rangle$, where $\langle \cdot \rangle$ denotes averages over trajectories of the particle with different initial
conditions and noise realizations.  We first consider the case
$\tau>\tau^*$. In Fig.~\ref{fig:tau10}(a) we show, as a function of
$F$, $\langle v_x \rangle$ and the mobility $\mu=\langle v_x
\rangle/F$ (inset), for $\tau=10$ and different values of $D_0$, as
computed in numerical simulations \cite{numsim}.  A linear regime at
small forces, characterized by a constant mobility depending on $D_0$, is followed by a complex nonlinear
scenario which emerges at intermediate values of the force.  In
particular, a nonmonotonic behavior corresponding to NDM takes place,
with a maximum that slightly shifts and then disappears as $D_0$ is
increased. This is expected because if the noise is strong enough the
effect of the velocity field $\boldsymbol{U}$ is averaged out. The
same happens for large enough forces, for which again the effect of
the velocity field $\boldsymbol{U}$ is negligible and, irrespective of
$D_0$, the trivial behavior $\langle v_x\rangle(F)=\tau F$ is recovered. Notice that
this asymptotic linear behavior is different from the saturation
effect at large force usually observed in lattice gas
models~\cite{BIOSV14}.

An even more striking phenomenon is observed in cases with $\tau\sim\tau^*$. In
Fig.~\ref{fig:tau10}(b) we show the force-velocity relation for
$\tau=1$: Again a complex nonlinear behavior can be observed for
intermediate values of $F$ and, surprisingly, ANM (i.e. $\langle v_x \rangle/F<0$) is observed in a range around $F\sim 0.065$. 

\begin{figure}[!t]
\includegraphics[width=.9\columnwidth,clip=true]{ballistic.eps}
\includegraphics[angle=-90,width=.95\columnwidth,clip=true]{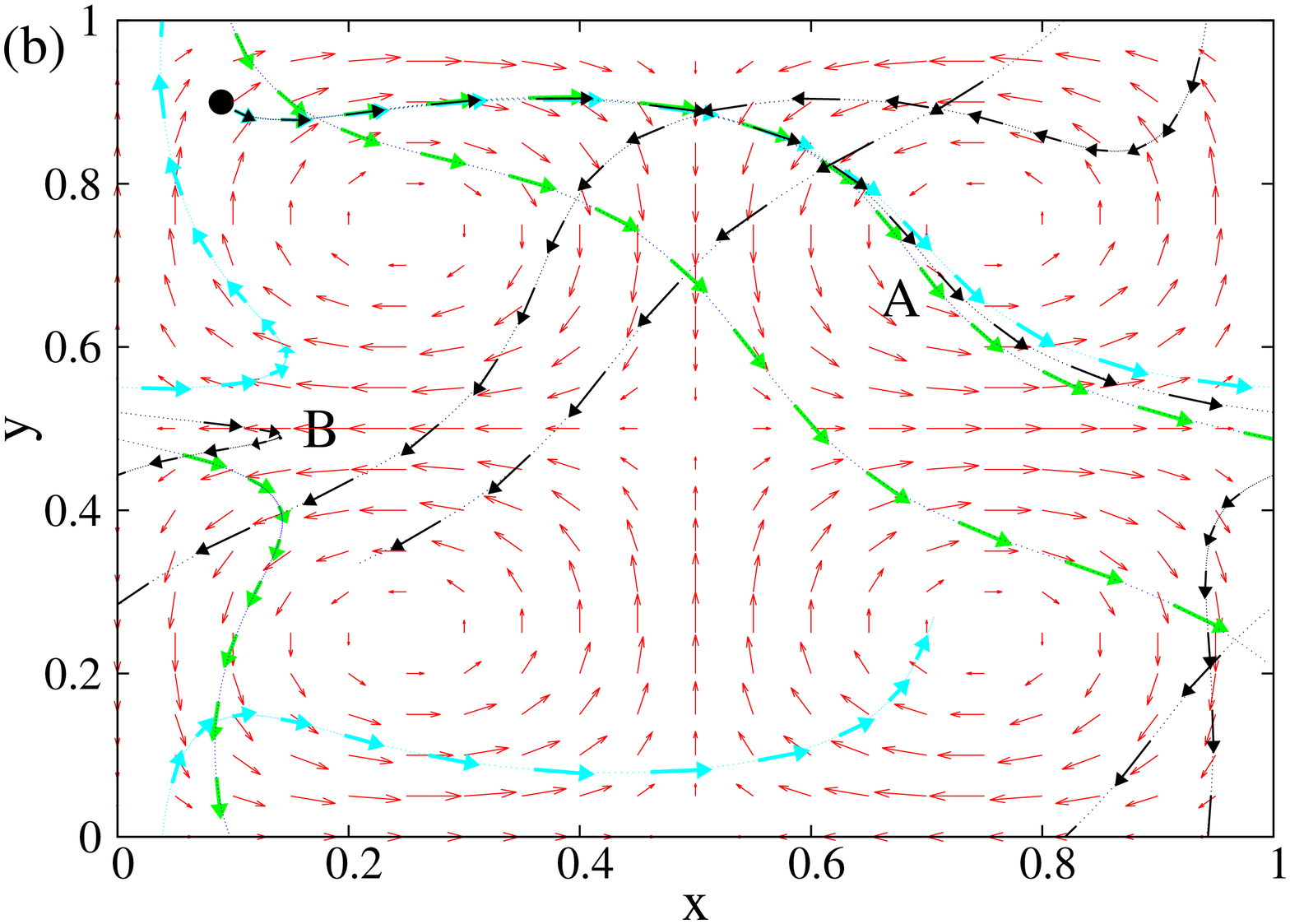}
\caption{Samples of the tracer's trajectories for $\tau=1$,
    $F=0.065$, $D_0=0$. (a) History of the particle's position
    $(x,y)$, recorded for a time-length $760 \tau^*$, starting near
    $(0,0)$ (marked as a black spot); (b) the black arrows represent the same
    trajectory for a timelength $\sim 10 \tau^*$ (folded into a single
    cell, for the purpose of visualization), the green and cyan arrows
    start with the same initial condition but are realized with
    $F=0.04$ and $F=0.09$ respectively, and the red arrows illustrate the
    underlying velocity field.}
\label{fig:traj}
\end{figure}

In order to get insight into the origin of the observed NDM and
  ANM, we have studied typical trajectories of the tracer as reported
  in Fig.~\ref{fig:traj}. In panel (a) we show that the motion of our
  tracers is realized along preferential ``channels'' which are
  aligned to two main directions: some of these channels are
  characterized by $\langle v_x \rangle<0$ (we call them ``leftward'')
  and others by $\langle v_x \rangle >0$ (called in the following
  ``rightward''). These preferential channels are seen for not too
  strong values of the noise (roughly up to values $D_0 \sim 10^{-4}$)
  and independently of the value of the force, but disappear reducing
  inertia. Both inertia and noise activate random transitions between
  the channels~\cite{MKTLH07}.  The force induces a bias in such
  transitions, determining, in general, an average $\langle v_x
  \rangle \neq 0$. In panel (b) we show a mechanism for explaining how
  an increase of the positive force may enhance the probability of
  transitions from rightward channels to leftward channels, which can
  lead to NDM or even ANM.  The reasoning is the following. Initially
  the particle is travelling along a rightward channel. For the chosen
  $F=0.065$ (black arrows), it occurs that the particle is pushed from
  region ``A'' to region ``B'' where the underlying velocity field is
  strongly negative: a transition to a leftward channel is then
  realized. With a smaller (green arrows) or larger (cyan arrows)
  force, the particle avoids the adverse region ``B'' and continues
  its run along rightward channels. This suggests that there exists a
  range of forces for which the tracer is induced to visit more
  frequently channels with velocity opposite to the force. Depending
  on how much this effect is pronounced, NDM or ANM can occur.

\begin{figure}[!t]
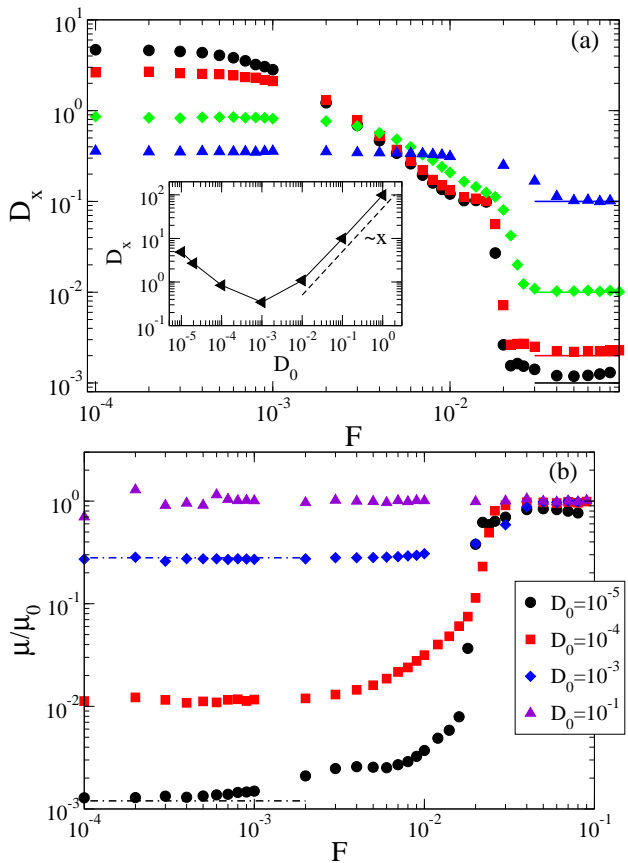

\includegraphics[width=.95\columnwidth,clip=true]{D_f.eps}
\includegraphics[width=.95\columnwidth,clip=true]{Teff2.eps}
\caption{(a) Tracer diffusivity $D_x(F)$ for different values of
 $D_0$ (see legend of Fig.~\ref{fig:tau10}(a)) with $\tau=10$. The continuous lines represent the values $\tau^2 D_0$. In
 the inset it is plotted $D_x(F=0)$ as a function of the microscopic
 diffusivity $D_0$. (b) Mobility $\mu$ over $\mu_0=D_x/T$, for $\tau=10$. The dot-dashed lines represent
  the predictions of the GER in two cases. }
\label{fig:maes}
\end{figure}

{\em Diffusivity}-- Next we focus on the study of the diffusion
coefficient $D_x(F)$, defined as
\begin{equation}
D_x=\lim_{t\to\infty}\frac{1}{2t}[\langle x(t)^2\rangle-\langle x(t)\rangle^2],
\end{equation}
in order to understand how the FDT is modified in our system.  Here we
consider the case $\tau=10$ (other cases show similar behaviors), which
is reported in Fig.~\ref{fig:maes}(a).  Notice that $D_x$ is nearly
independent of the force at small forces and at large forces, where it
coincides with the value expected in the absence of the velocity
field, $D_x=\tau^2 D_0$. It is remarkable that $\lim_{F \to 0} D_x \gg
\lim_{F \to \infty} D_x$: such a discrepancy decreases when $D_0$ is
increased.  In order to better understand the role of the molecular
diffusivity $D_0$ in our system, in the inset of Fig.~\ref{fig:maes}(a)
we report the behavior of $D_x (F=0)$ as a function of $D_0$.  For
large enough noise amplitude, the scaling is linear, as expected,
because the diffusion coefficient is dominated by the microscopic
diffusivity.  On the contrary, for $D_0 \to 0$, the particle
diffusivity diverges, similarly to what found by Taylor~\cite{Taylor}
for the dispersion of a fluid particle in laminar flows in straight
channels. In the case of Taylor diffusion of a fluid particle in a
shear flow, the behavior $D_x\sim D_0^{-1}$ can be easily understood
in terms of long horizontal ballistic motion, the duration of which
increases as $D_0$ decreases. In our system the understanding is
  not so simple, but the divergence $D_x\sim D_0^{-\alpha}$ and the
  long channels observed in Fig.~\ref{fig:traj}(a)
  suggest a similar scenario.

\emph{Generalized Einstein relation.}-- The behaviors described
above can be interpreted within the context of response theory.  In
equilibrium conditions, and in the linear regime, the Einstein
relation predicts a proportionality between the mobility and the
diffusivity, via the inverse temperature
\begin{equation}
\mu=\mu_0\equiv\frac{1}{T}D_x.
\label{ER}
\end{equation}
In our system, the presence of the velocity field $\boldsymbol{U}$
introduces significant nonequilbirium effects that are clearly visible
in Fig.~\ref{fig:maes}(b), where we report the measured mobility $\mu$
rescaled by $\mu_0$. Only for large enough values of $F$, where the
effect of $\boldsymbol{U}$ is negligible and the system can be
considered at equilibrium, the ratio $\mu/\mu_0\sim 1$. Eventually, for $D_0$ large enough,
the noise makes the velocity field irrelevant and $\mu/\mu_0\sim 1$ in all regimes.

The difference due to nonequilibrium effects can be revisited in
terms of a GER, derived for systems in out-of-equilibrium steady
states. According to this relation, the particle mobility can be
expressed as the sum of two contributions: one proportional to the
diffusion coefficient, as in the standard Einstein
relation~(\ref{ER}), and the other involving the correlation function
with the time-integral of the velocity field $U_x$, computed along the
trajectory of the particle. As discussed in detail in~\cite{BMW11},
for a system described by a set of stochastic equations as in
Eqs.~(\ref{eq1}-\ref{eq4}), the GER explicitly reads
\begin{align}
\lim_{F \to 0} \mu(F) = \frac{1}{T}[D_x(F=0) - C_{x\Phi}(F=0)] \\
C_{x\Phi}(F) = \lim_{t\to\infty}\frac{1}{2Tt}\langle [x(t)-x(0)]\Phi(t)\rangle_{c,F},
\label{ger}
\end{align}
where
\begin{equation}
\Phi(t)=\int_0^t U_x[x(s),y(s)]ds,
\end{equation}
and $\langle A(t)B(s)\rangle_{c,F}=\langle A(t)B(s)\rangle_{F}-\langle
A\rangle_{F} \langle B\rangle_{F}$ is the connected correlation
function measured at force $F$.  We have computed in numerical
simulations the nonequilibrium contribution due to the coupling with
the field $\Phi$. The validity of the predictions of the
GER~(\ref{ger}) is shown - as dot-dashed lines - in
Fig.~\ref{fig:maes}(b), for two cases at $D_0=10^{-5}$ and
$D_0=10^{-3}$.  Let us stress that Eq.~\eqref{ger} can be
  exploited also at non-vanishing forces: indeed the {\em
  differential} mobility $d\langle v_x \rangle/dF$ at a finite value
of $F$ is given by the same expression, by measuring the two terms
$D_x$ and $C_{x\Phi}$ at force $F$. The prediction of GER for
$d\langle v_x \rangle/dF$ is negative where NDM appears, as we checked
numerically: NDM therefore can be interpreted as the consequence of
$C_{x\Phi}$ becoming larger than $D_x$~\cite{BM14}.
Also in the case $\tau=1$, for the force values with ANM, the
GER is verified, showing strong negative and positive differential
mobilities just before and just after the minimum of $\langle v_x
\rangle (F)$. 

\emph{Conclusions.}-- We have studied the effects of a driving
external force on the dynamics of an inertial particle advected by a
velocity field, in the nonlinear regime. We have discovered nontrivial
behaviors of the stationary tracer velocity and of its diffusivity as
a function of the force, such as NDM and ANM. These effects are due to a
complicated combined action of the applied force, the particle inertia
and the underlying velocity field. It turns out that, in some force
regimes, this coupling leads the tracer to persist in regions of the
velocity field which drag it against the force direction, resulting in
a slowing down of the tracer velocity, or even producing a negative
mobility~\cite{nota1}. The central role played by the coupling with the velocity
field clearly emerges in the GER which is satisfied in our
nonequilibrium system.  The striking behaviors shown by the model
should be observable in experiments with biased inertial tracers in
laminar flows, realized, for instance, in setups with rotating
cylinders \cite{SG88}, two-sided lid-driven cavities \cite{KWR97} or
magnetically-driven vortices \cite{T02,PS05}.

\begin{acknowledgements}
We thank M. Cencini for useful discussions.
\end{acknowledgements}


\begin{thebibliography}{99}
\bibitem{nota0}
Note that $D_0$ refers to velocity (and not spatial) diffusion.
\bibitem{nota}
The limit $\tau \to 0$ has been also considered in~\cite{BG90} and references therein: 
there an anomalous diffusion has been observed, but in a transient regime only.
\bibitem{numsim}
The integration of the stochastic equations of the model is performed with a fourth-order Runge-Kutta
algorithm~\cite{H92}, with a time step $dt=10^{-4}$. Numerical results shown in the 
figures are averaged over more than $10^{4}$ realizations, and error bars fall within the symbols when not explicitly marked.
\bibitem{nota1} We note that the presence of persistent ballistic
  trajectories can be suppressed in non-laminar flows. For instance,
  we verified the absence of NDM in the case of a synthetic random
  field as in~\cite{B05}.
\end{thebibliography}
\end{document}